\documentclass[structabstract]{aa}  
\usepackage{natbib}\usepackage{txfonts}\usepackage{balance}
\usepackage{graphicx}
\usepackage{url}
\usepackage{txfonts}
\def\lax{\lesssim}
\def\gax{\gtrsim}
\def\gmrt{GMRT\,}

\def\chandra    {\emph{Chandra}\,}

\def\vla        {VLA\,}

\def\ie{i.e.\,}
\def\eg{e.g.\,}

\def \deg      {$^{\circ}$}
\def\arcsec{$^{\prime\prime}$}
\def\arcmin{$^{\prime}$}

\begin{document}
\title{153 MHz \gmrt follow--up of steep--spectrum diffuse emission in galaxy clusters}

   \author{
   G.~Macario\inst{1,}\inst{2}, 
   T.~Venturi\inst{2}, 
H.~T.~Intema\inst{3,}\footnote{Jansky Fellow of the National Radio Astronomy Observatory},
D.~Dallacasa\inst{2,}\inst{4},
G.~Brunetti\inst{2},
R.~Cassano\inst{2}, 
S.~Giacintucci\inst{5,}\inst{6,}\inst{2}, 
C.~Ferrari\inst{1}, 
C.~H.~ Ishwara-Chandra ~\inst{7},
R.~Athreya\inst{8} 
}

\institute
{
Laboratoire Lagrange, UMR7293, Universit\'e de Nice Sophia-Antipolis, CNRS, Observatoire de la C\^ote dÕAzur, 06300 Nice, France
\and
INAF -- Istituto di Radioastronomia, via Gobetti 101, I-40129, Bologna, Italy
\and
National Radio Astronomy Observatory, 520 Edgemont Road, Charlottesville, VA 22903-2475, USA
\and
Dipartimento di Astronomia, Universit\'a di Bologna, via Ranzani 1, I--40127,
Bologna, Italy
\and
Department of Astronomy, University of Maryland, College Park, MD 20742-2421
\and 
Joint Space-Science Institute, University of Maryland, College Park,
MD, 20742-2421, USA
\and
National Centre for Radio Astrophysics, TIFR, Ganeshkhind, Pune 411007, India
\and
Indian Institute of Science Education and Research, Sutarwadi Road, Pashan, 
Pune 411021, India
}

\date{Received 00 - 00 - 0000; Accepted 00 - 00 - 0000}

\titlerunning{153 MHz \gmrt follow--up of steep--spectrum cluster diffuse emission}

\authorrunning{Macario et al.}

\abstract
{}
{In this paper we present new high sensitivity 153 MHz Giant Meterwave Radio 
Telescope follow--up observations of the diffuse steep spectrum cluster radio 
sources in the galaxy clusters Abell\,521, Abell\,697, Abell\,1682. 
Abell\,521 hosts a relic, and together with Abell\,697 it also hosts  a giant 
very steep spectrum radio halo. Abell\,1682 is a more complex system with 
candidate steep spectrum diffuse emission.}
{
We imaged the diffuse radio emission in these clusters at 153 MHz, 
and provided flux density measurements of all the sources at this frequency.   
Our new flux density measurements, coupled with the existing data at
higher frequencies, allow us to study the total spectrum of the halos and
relic over at least one order of magnitude in frequency.}
{Our images confirm the presence of a very steep 
\textquotedblleft{diffuse component}\textquotedblright\ in Abell\,1682.
We found that the spectrum of the relic in Abell\,521 can be fitted by a single 
power--law with $\alpha=1.45\pm0.02$ from 153 MHz to 5 GHz.
Moreover, we confirm that the halos in Abell\,521 and Abell\,697 have a very
steep spectrum , with $\alpha=1.8-1.9$ and $\alpha=1.52\pm0.05$ respectively.
Even with the inclusion of the 153 MHz flux density information it is 
impossible to discriminate between power--law and curved spectra, as
derived from homogeneous turbulent re--acceleration. The latter are favored
on the basis of simple energetic arguments, and we expect that LOFAR will
finally unveil the shape of the spectra of radio halos below 100 MHz,
thus providing clues on their origin.  
} 
{}
   \keywords
   {radiation mechanism: non-thermal -- galaxies: clusters: general -- 
galaxies: clusters: individual: A\,521, A\,697, A\,1682}

   \maketitle
   

\section{Introduction}

Clusters of galaxies are the largest and most massive gravitationally
bound systems in the Universe, with typical total mass content up to a
few 10$^{15}$ M$_{\odot}$ in the form of galaxies, hot intra--cluster medium
(ICM) and dark matter in order of increasing fraction. \\
The existence of non-thermal components (cosmic rays and magnetic fields) 
on the cluster scale permeating the ICM 
is nowadays well established thanks to deep radio observations.  
These show the existence of diffuse extended (up to and above Mpc) 
radio synchrotron sources with no obvious optical counterpart, in 
$\sim$50 massive galaxy clusters. 
According to their morphology and polarization properties, 
diffuse Mpc-scale radio sources are classified as halos or relics 
(\eg, Ferrari et al. 2008; Venturi 2011, for recent reviews).
They probe the presence of $\sim\mu$G large scale magnetic fields and 
relativistic ($\sim$GeV) electrons in the ICM.\\ 
Radio halos and relics share some observational properties:  
they both have steep synchrotron spectra, with typical spectral index values
$\alpha\sim$1.2--1.4 (in the convention S$\propto \nu^{-\alpha}$), 
and are characterized by very low surface brightness emission
($\sim\mu$Jy/arcsec$^2$). 
However, they differ in the location within galaxy clusters,
as well as in their polarization properties. Halos are centrally located,
their size and shape are usually similar to the
distribution of the X--ray brightness coming from the intra--cluster gas, 
and they are unpolarised (the only confirmed exception being 
MACS\,J0717+3745, Bonafede et al. 2009). 
Relics are found in the outskirts of galaxy clusters, and in most cases
are found at the border of the X--ray emission from the ICM; 
they exhibit a variety of shapes and have high fractional polarization.

There is now compelling observational evidence in favor of an
unambiguous connection between cluster mergers and the detection of radio
halos and/or relics. In particular, a quantitative  radio/X--ray analysis of 
the GMRT (Giant Metrewave Radio Telescope)
cluster radio halo sample (Venturi et al. 2007 and 2008, hereinafter V08) 
carried out in Cassano et al. (2010), shows that radio halos are always 
found in clusters with high disturbance, as derived by a number of 
X--ray morphological indicators, 
while relaxed systems never host one. A few outliers exist, \ie clusters
dynamically disturbed but with no radio halo detected at the sensitivity 
levels of the current instruments.
The spectral properties of some radio halos (\eg, Schlickeiser et al. 1987;
Thierbach et al. 2003; Brunetti et al. 2008; Donnert et al. 2010), combined
with the radio halo--cluster merger connection (Cassano et al. 2010,
Brunetti et al. 2007) support a scenario based on turbulent re--acceleration
for radio halos, whereby the emitting electrons are re--accelerated by
turbulence in the intra--cluster medium during cluster mergers (Brunetti et 
al. 2001; Petrosian 2001). An alternative possibility to explain the origin
of radio halo is provided by the ``secondary model'', where relativistic
electrons are the secondary products of inelastic proton--proton collisions 
in the ICM (\eg Dennison 1980; Blasi \& Colafrancesco, 1999).
\\
Complementary, the origin of relics seems to be related to the presence of 
accreting and merger shocks in the outer regions of the clusters, even though 
a clear connection between relics and shocks has been so far circumstantial
(Abell\,521, Giacintucci et al. 2008, hereafter G08; Abell\,754, Macario et al. 2011; 
RXC\,J1314.4--2515, Mazzotta et al. 2011; likely CIZA\,J2242.8+5301, see Ogrean et al. 2013).

Despite their steep spectra, halos and relics have been mainly
imaged at GHz frequencies. Only recently, thanks to the very good imaging
capabilities of the GMRT at low frequencies and to the advent of LOFAR 
(LOw Frequency ARray), it has become possible to perform high sensitivity 
observations of halos and relics at sub--GHz frequencies down to 150 MHz 
and below,
thus allowing us to probe the \textquotedblleft{low end}\textquotedblright\, 
of the energy distribution of the relativistic electron population 
(\eg van Weeren et al. 2011, Kale \& Dwarakanath 2009 and 2010, van Weeren et al. 2012). 
This is essential to understand the origin of these sources, since the
acceleration mechanisms at play have a clear imprint on the particle
spectrum (and hence on the radio emission).
To date, accurate flux density measurements of radio halos and relics below 
$\nu \lax$ 300 MHz are known only for a handful of sources, and
this field of investigation is still largely unexplored. 

The low--frequency follow--up of the GMRT Radio Halo sample (Venturi et al. 
2009, Giacintucci et al. 2011, Venturi et al. 2013, hereinafter V13) led to the 
unexpected discovery of a family of radio halos with very steep spectrum, \ie  
$\alpha \sim 1.5 \div 2$, barely visible at GHz frequencies but clearly
detected and imaged at $\lesssim$ 325 MHz. The prototype of this class of 
sources was found in Abell\,521 (Brunetti et al. 2008, hereinafter B08; 
Dallacasa et al. 2009, hereinafter D09). Radio halos with ultra steep
spectrum are extreme cases  of this class of cluster sources. Their very steep
spectrum suggests that the re--acceleration mechanisms are poorly efficient, 
since the energy of the re--acceleration is concentrated in the 
lowest energy electrons. At the same time, the very steep spectrum of these
sources challenges some of the proposed models for their origin, such as 
the secondary models, which would require an unplausibly high energy budget
(Brunetti 2004; B08).

With the aim to investigate the properties of diffuse cluster sources at 
very low frequency, we performed GMRT observations at 153 MHz of three
clusters: Abell\,521 and Abell\,697, both hosting a very steep spectrum radio halo, 
(B08; D09; Macario et al. 2010, hereinafter M10), and Abell\,1682. 
The latter is characterized by extended radio emission on the cluster scale, 
which is difficult to fit into the halo/relic classification (V08, Venturi et al. 2011, hereinafter V11).
The present 153 MHz observations, combined with the radio data already
available in the literature, allow us to study the spectra of the diffuse 
cluster sources in Abell\,521 and Abell\,697 with five data points in the frequency 
range 153 MHz -- 1.4 GHz, and to further investigate the nature of the 
emission in Abell\,1682.

The paper is organized as follows: in Sect. \ref{sec:sample} we summarize 
the properties and literature information of the three clusters.  
In Sect. \ref{sec:obs} we present the new \gmrt radio observations and data 
reduction procedure. 
Radio images are presented in Sect. \ref{sec:images}; the data analysis and 
spectral results are described in Sect. \ref{sec:analysis}.
In Sect. \ref{sec:discussion} we discuss our results in the light of the  
current models for the origin of radio halos and relics, with particular
attention to their spectral properties at low frequency.
A summary and concluding remarks are given in Sect. \ref{sec:summary}. \\
Through the paper, we adopt a $\Lambda$CDM cosmology, with 
H$_0$=70 km s$^{-1}$ Mpc$^{-1}$, $\Omega_m=0.3$ and $\Omega_{\Lambda}=0.7$. 
Hereafter, we will use the abbreviation \textquotedblleft{A}\textquotedblright\, 
for \textquotedblleft{Abell}\textquotedblright\, clusters. 


\section{The cluster sample}
\label{sec:sample}

\subsection{A\,521: the prototypical USSRH}
\label{sec:sam521}

A\,521 is a massive and X--ray luminous merging cluster
(M$_V \sim 1.9 \times 10^{15}$ M$_{\odot}$; 
L$_{X[0.1-2.4 \rm keV]} \simeq 8 \times 10^{44}$ erg s$^{-1}$), 
located at redshift z=0.247 (1\arcmin = 232.2 kpc). 
Multi-wavelegth studies show that this system is undergoing multiple merging 
episodes between smaller sub--clusters (\eg Ferrari et al. 2003). 
It hosts both a peripheral relic and a central giant radio halo, which have 
been studied in detail at various frequencies (Ferrari et al. 2006; 
Giacintucci et al. \ 2006, G08, B08 and D09). 

The relic source is detected up to 5 GHz, and its spectrum in 
the range 240 MHz--5 GHz is well fitted by a single power--law with spectral 
index $\alpha\simeq 1.5$ (G08, B08). It is one of the very few cases with 
a detection of a X--ray front coincident with the relic emission 
(G08, Bourdin et al. 2012).
The central radio halo is barely detected at frequencies $\gax$ 610 MHz, 
and becomes clearly visible only below 330 MHz (G08, B08). 
Its integrated spectrum is very steep, with $\alpha \simeq 1.9$ over the range 
240 MHz--1.4 GHz (D09 and references therein), thus it is considered 
the prototypical USSRH.  

\subsection{A\,697}
\label{sec:sam697}

A\,697 is a hot ($k$T $\simeq$ 10 keV), luminous 
(L$_{X}\simeq10^{45}$ erg s$^{-1}$) and massive cluster 
(M$_V \sim 2.3 \times 10^{15}$ M$_{\odot}$) 
at z=0.282 (1\arcmin = 255.6  kpc). 
Observational evidence suggests that A\,697 is in a complex dynamical state,
and it is most likely undergoing multiple merger/accretion of small clumps: 
substructure in the galaxy distribution and in the gas has 
been detected through optical and X-ray analysis (\citealt{girardi06}; M10). 
\\
The giant radio halo in A\,697 was clearly detected with the \gmrt at 610 MHz, 
as part of the \gmrt radio halo survey observations (V08). 
In M10, using deep 325 MHz \gmrt observations together with the 610 MHz 
\gmrt data and VLA archival data at 1.4 GHz, we showed that the integrated 
radio spectrum of the halo is very steep, with 
$\alpha^{1.4 \rm{GHz}}_{325 \rm{MHz}} \approx 1.7-1.8$. \\
The high frequency end of the spectrum of the halo was recently constrained by 
WSRT observations at 1.4 and 1.7 GHz  (Van Weeren et al. 2011), who fit the 
spectrum with a single power--law with 
$\alpha^{1.7 \rm GHz}_{325 \rm MHz}=1.64\pm0.06$. This value, though slightly 
flatter, is consistent with our earlier result. 

\subsection{A\,1682: a very complex cluster}
\label{sec:sam1682}

A\,1682 is an X-ray luminous (L$_{X[0.1-2.4 keV]} \simeq 7 \times 10^{44}$ erg s$^{-1}$) 
cluster located at redshift z = 0.2260 (1\arcmin = 237.2 kpc). 
X--ray and optical observations suggest the cluster has experienced a recent 
merger (\eg Dahle et al. 2002, Morrison et al. 2003). 

A\,1682 shows a very complex radio morphology at 610 MHz (V08). 
The radio emission is dominated by an extended tail at the cluster centre 
(named \emph{E-tail}), and by two features referred to as \emph{S--E }and 
\emph{N--W ridge} (see  Fig.  6 in V08), whose nature is still unclear. 
Beyond those individual sources, V08 reported on the presence of positive 
residuals, hinting at the existence of low surface brightness diffuse 
emission, embedded in the emission from the main radio sources and extended 
on the cluster scale. 
\gmrt follow--up observations at 240 MHz clearly revealed the presence of 
another feature, named 
\textquotedblleft{diffuse component} \textquotedblright\, 
(V11 \& V13). 
This has a strong counterpart in the VLSS (Very Low frequency Sky Survey, 
Cohen et al. 2007) image, and has been imaged with the VLA in the A 
configuration at 74 MHz  (see  Fig.  1 of V11, left panel;
Dallacasa et al. in prep.). 
Those observations suggest that the 
\textquotedblleft{diffuse component} \textquotedblright\, 
has a very steep spectrum.

\section{Observations and data reduction}
\label{sec:obs}

In Table \ref{tab:obs} we report the main details of the \gmrt 153~MHz 
follow--up observations of our cluster sample. All clusters were observed 
in August 2009. 
In order to achieve high sensitivity and to ensure good u--v coverage, each 
cluster was observed for a total time of $\sim$10 hours. To minimize radio 
frequency interference (RFI) and scintillation, night--time 
observations were carried out for all the clusters, except for A\,1682, due 
to scheduling constrains. The daytime observations affected significantly our 
results, as discussed later in this section. 
The observations were recorded in one sideband 8~MHz wide, split into 
128 spectral channels of 62.5~kHz each.

\begin{table*}[ht!]
\caption{Summary of \gmrt 153 MHz observations}
\begin{center}
\begin{tabular}{lccccccccc}
\hline
Cluster & RA$_{J2000}$ & DEC$_{J2000}$ & z & Obs. Date &$\nu$ ($\Delta\nu$)  & $t_{\rm obs}^{\rm (a)}$ & $t_{\rm eff}^{\rm (b)}$ &HPBW, PA& rms$^{\rm (c)}$ \\
 name			      & (h,m,s \& \deg, \arcmin, \arcsec)			    & 			     &    &                  &(MHz)        & (hours) & (hours)&(\arcsec $\times$ \arcsec, \deg)
			      &(mJy beam$^{-1}$)\\
\hline
A\,521  	& 04 54 09.1 &  -10 14 19  & 0.2475  &2009,  Aug 16  & 153  (8) & 10& 5.1  & 34.7$\times$20.9, 59.0& 0.9\\
A\,697 	& 08 42 53.3 & +36 20 12  & 0.2820  & 2009, Aug 30 &  153 (8) & 10& 6.9 & 26.2$\times$20.8, -89.3& 0.8\\
A\,1682  & 13 06 49.7 & +46 32 59 & 0.2260  & 2009, Aug 17 & 153  (8) & 9& 5.6 & 29.5$\times$17.9, 53.8& 1.7 \\
\hline
\end{tabular}
\end{center}
$^{\rm (a)}$ Total duration of the observation. \\
$^{\rm (b)}$ Effective time on source. \\
$^{\rm (c)}$ Measured in the central portion of the image. 
\label{tab:obs}
\end{table*}

The data reduction was performed using several software packages, namely AIPS 
\citep[e.g.,][]{bridle1994}, Obit \citep{cotton2008}, ParselTongue 
\citep{kettenis2006} and SPAM \citep{intema2009}. 
AIPS provided us with the basic functionality needed for the largest part 
of the data reduction, accessed from the Python programming language using 
the ParselTongue interface. 
Obit was used for RFI excision (see below), and SPAM was used to derive 
and apply direction--dependent calibrations during (AIPS-based) imaging. 
\\
For each observation, the flux scale and bandpass shapes were derived from 
a $\sim 20$~minute observation on 3C\,147, adopting a flux of 60.0~Jy at 
153~MHz (for a discussion on calibrator flux scales below 330~MHz, 
see \citealt{intema2011}; see also \citealt{scaife2012}, for a more recent study of low frequency calibrators spectral models). 
Furthermore, 3C\,147 was used to estimate the instrumental phase contribution 
to the antenna gains, needed for SPAM ionospheric calibration \citep[for 
details, see][]{intema2009}. The bandpass, flux scale and instrumental phase 
information were used to correct the target field data, resulting in an 
effective bandwidth of 6.875~MHz.

Strong RFI and other bad data were removed from the target field data  after 
visual inspection, simple clipping bad visibility amplitudes, and 
subtracting persistent RFI \citep{athreya2009}. 
As next step, to speed up the processing, the data volume was averaged in 
frequency to 22~channels of 312.5~kHz each, to avoid significant bandwidth 
smearing. We phase--calibrated each target field using a 10~component point 
source model derived from the NRAO VLA Sky Survey (NVSS, Condon et al. 1998) 
and from the WEsterbork Northern Sky Survey (WENSS, Rengelink et al. 1997), 
followed by wide-field imaging and cleaning of the full field-of-view with 
75 facets. 
Rounds of self-calibration and wide-field imaging and inspection of the 
residual visibility data for further editing were repeated several 
times, with amplitude calibration in the final round only. 

To address the deconvolution problems due to direction--dependent ionospheric 
phase errors, we applied SPAM calibration and imaging to the target fields. 
An essential step in this approach is to individually calibrate on several 
bright sources within the field-of-view, a technique known as \emph{peeling} 
\citep[e.g.,][]{noordam2004}. If enough calibrator sources are available,
the phases can be fitted with an ionosphere model to predict phase corrections 
in arbitrary viewing directions. This approach worked well for A\,521 and 
A\,697, but failed for A\,1682 due to poor data quality. 
For A\,521 and A\,697 the ionospheric calibration model was applied during 
imaging and cleaning of all facets.
In the case of A\,1682, we applied the peeling phases towards the few 
calibrators, but used the (non--directional) self--calibration for all other 
facets. The SPAM calibration and imaging were repeated twice per target.
The final images of the full (3~degree) field-of-view centered on our 
clusters were corrected for primary beam attenuation. 
Details of our final images are given in Table \ref{tab:obs}.\\

At the end of the data reduction process, the total data loss was $\sim~45$\% 
for A\,521 and A\,697, nevertheless, the quality of the final images met our 
requirements.
The higher data loss of ~57\% for A1682 was caused by several power 
failures of the array, malfunctioning antennas, and more severe RFI and 
ionospheric scintillations due to daytime observing. 
The noise level in the final A1682 image is a factor of $\sim$2 worse than 
for A521 and A697.  
Residual RFI and ionospheric effects are the most likely cause.\\
We estimate that the residual amplitude errors are of the order of 
$\lax$15 \%,  in  line with values reported for \gmrt observations at 
this frequency (see \eg Intema et al. 2011; Sirothia et al. 2009; 
Kale \& Dwarakanath 2010). 
However, there are several factors that may affect the derived flux scale 
from the primary calibrator \citep[e.g., see][]{intema2011}. 
As a final step we thus checked for systematic offsets in the flux scale in 
each field. To this aim, we selected a number ($\sim 30$) of bright and compact 
sources in each field, and  compared the measured GMRT 153 MHz flux density 
with a model spectrum flux, derived using catalog fluxes at 1.4 GHz (NVSS), 
325 MHz (WENSS) and at 74 MHz (VLSS, Cohen et al. 2007). 
We concluded that the uncertainty in the flux 
density scale is of the order of 20\% in each field.

For convenience, in order to image only the inner portion of each field 
without the need of wide-field imaging technique and 
focus on the central emission in our clusters, from the latest uv--data we subtracted 
all the discrete radio sources in the 3 degrees field, except those in the central 
area of $\sim0.3$ degrees. 
These \textquotedblleft{central}\textquotedblright\ datasets were 
used to produce the final images useful to our analysis (Section \ref{sec:images}). 

\section{153 MHz images}
\label{sec:images}

   \begin{figure*}[Ht!]
   \centering
\includegraphics[angle=0,scale=0.53]{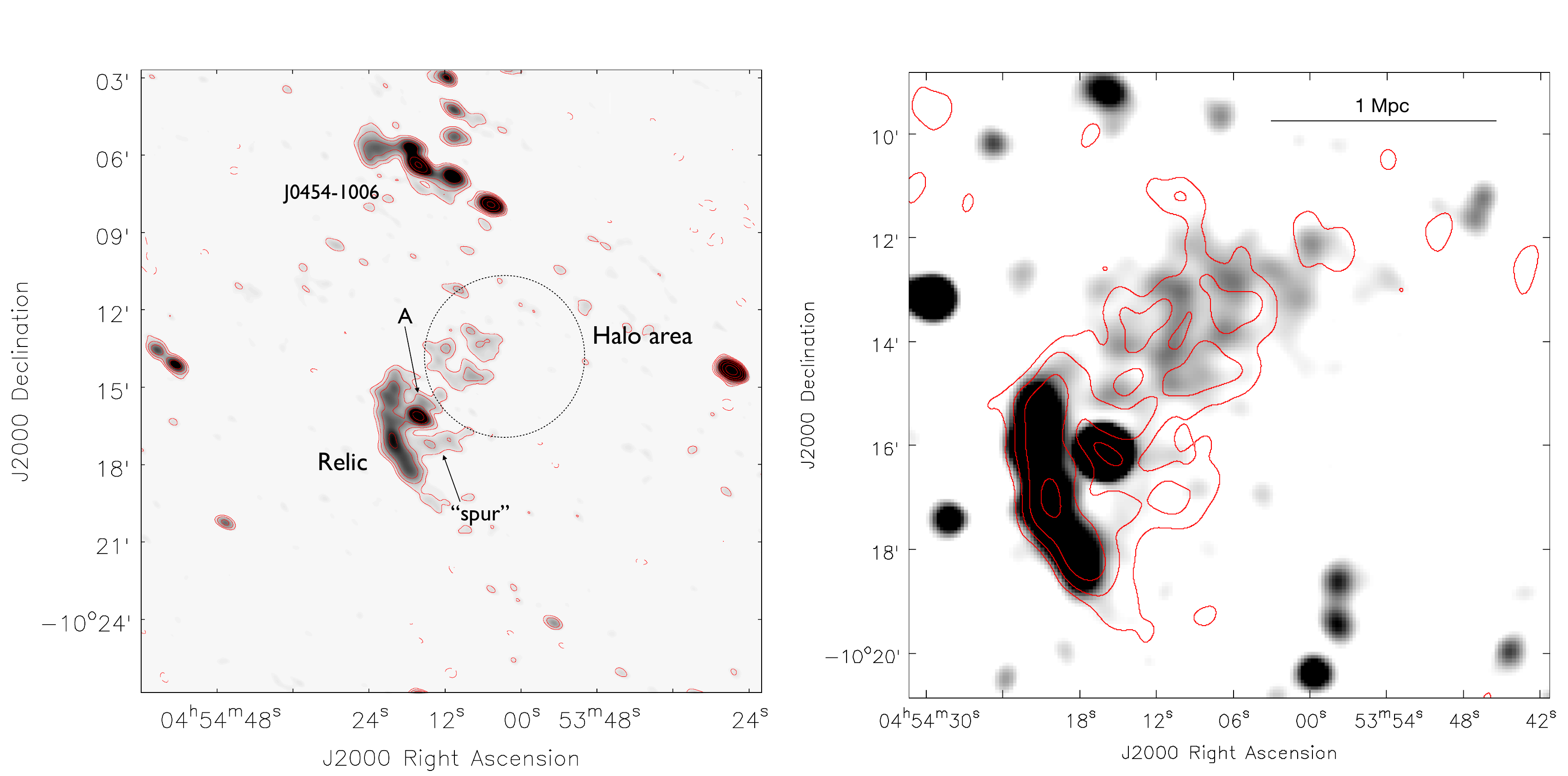}
      \caption{ 
{\it Left} -- \gmrt 153 MHz contours of the central field around the cluster 
A\,521. Contours start at $\pm$3$\sigma_{153 \rm MHz}$=$\pm$2.7 mJy/b 
and are spaced by a factor of two. 
The resolution is $34.7^{\prime\prime}\times20.9^{\prime\prime}$, p.a. 59.0\deg. 
The dashed circle represent the area of $~$1 Mpc containing the central radio 
halo (same as in  Fig.  1 of B08). 
{\it Right} -- \gmrt 153 MHz contours of the central diffuse emission in 
A\,521 (halo, relic), after the subtraction of source A.  
Contours start at $\pm$3$\sigma_{153 \rm MHz}$= $\pm$4.35 mJy/b and are spaced 
by a factor of two, overlaid to the VLA 1400 MHz image, in greyscale (from 
D09; the lowest level of greyscale corresponds to the 
3$\sigma_{1400 \rm MHz}$=90$\mu$Jy/b, the highest correspond to 0.55 mJy/b, and 
the peak flux density is  15.5  mJy/b). 
The two images have similar angular resolutions:  
$38.0^{\prime\prime}\times35.0^{\prime\prime}$, p.a. 0\deg at 153 MHz and 
$30.0^{\prime\prime}\times30.0^{\prime\prime}$, p.a. 0\deg at 1400 MHz. 
}
\label{fig:fig1}
   \end{figure*}


   \begin{figure*}
   \centering
\includegraphics[angle=0,scale=0.5]{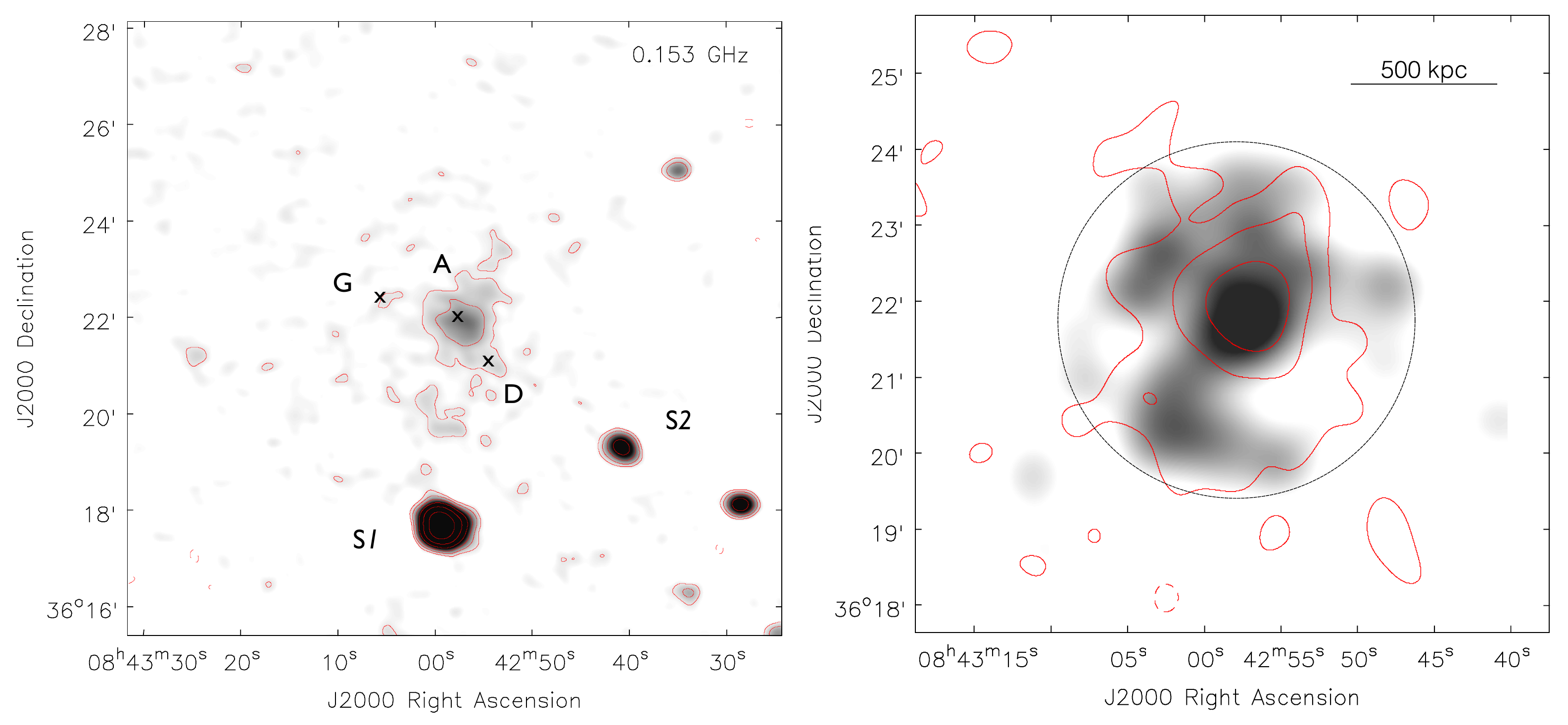}
      \caption{
	\textit{Left}--\gmrt 153 MHz full resolution image of A\,697. 
Contours are spaced by a factor of two, starting from 
$\pm$3$\sigma_{153 \rm MHz}$= $\pm$2.4 mJy/b. The restoring beam is 
$26.2^{\prime\prime}\times 20.8^{\prime\prime}$, p.a. -89.5\deg. 
Crosses and labels mark the position of the point sources detected and 
optically identified at 325 MHz (M10). 
	\textit{Right}--\gmrt 153 MHz low resolution contours of the central 
radio halo in A\,697, starting at $\pm$3$\sigma_{153 \rm MHz}$= $\pm$3.6 mJy/b 
and spaced by a factor of two. The resolution is 
$48.4^{\prime\prime}\times44.4^{\prime\prime}$, p.a. 3.5\deg. Contours are overlaid 
on the 325 MHz image (shown in greyscale, lowest level corresponding to 
3$\sigma_{325 \rm MHz}$), at a resolution of 
$46.8^{\prime\prime}\times41.4^{\prime\prime}$, p.a. 79.7\deg (M10).  
              }
\label{fig:fig3}
   \end{figure*}

  \begin{figure*}
   \centering
\includegraphics[angle=0,scale=0.5]{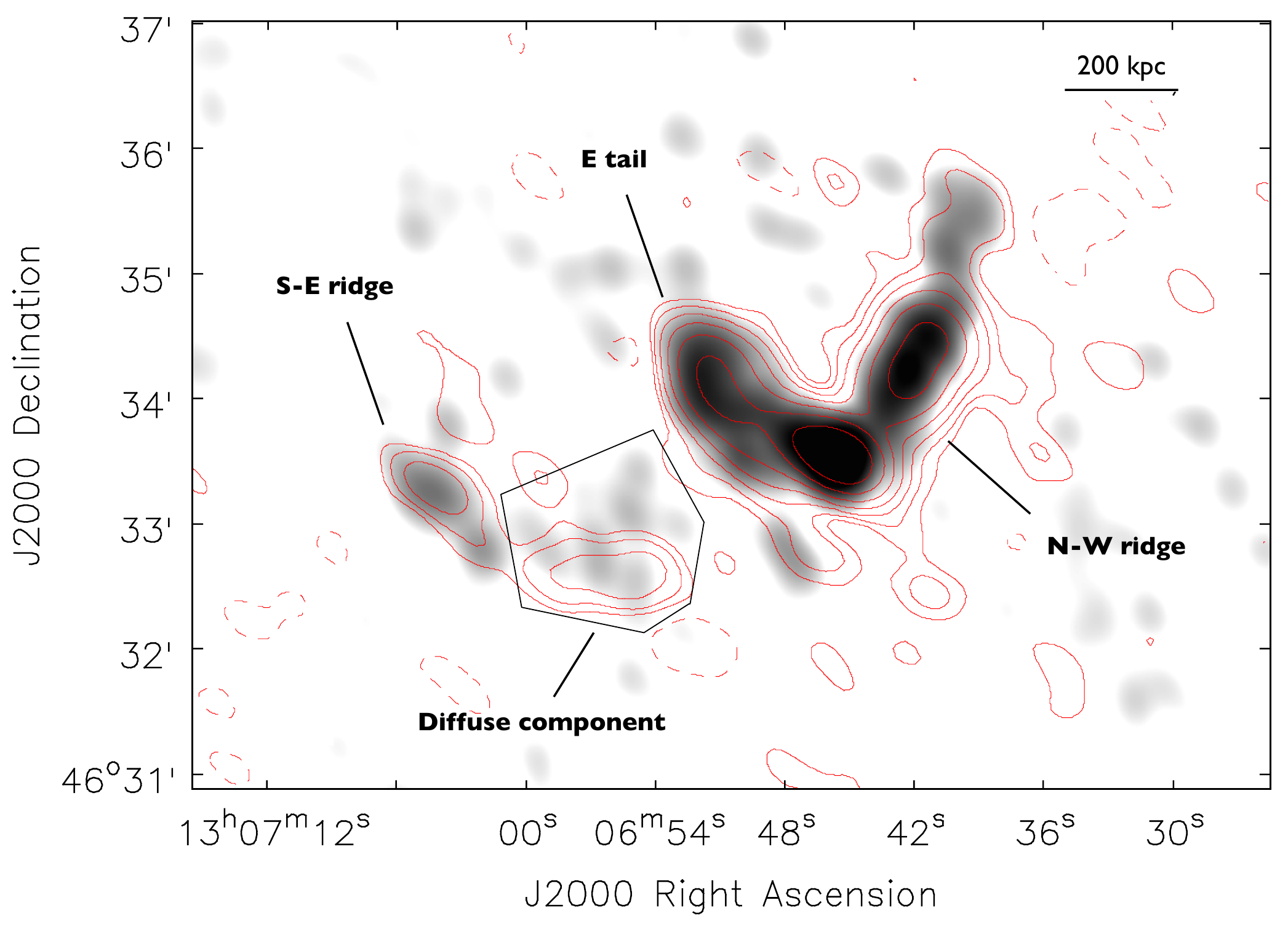}
      \caption{\gmrt 153 MHz contours of the central emission in A\,1682, 
starting at $\pm$3$\sigma_{153 \rm MHz}$= $\pm$5.1 mJy/b and spaced by a factor 
of two.  
The resolution is $29.5^{\prime\prime}\times18.0^{\prime\prime}$, p.a. 53.8\deg. 
Contours are overlaid to the 240 MHz image, shown in greyscale,   
at resolution $18.3^{\prime\prime}\times14.0^{\prime\prime}$, p.a. 21.7\deg 
(Venturi et al. 2009). The lowest level of the greyscale corresponds to 
3$\sigma_{240 \rm MHz}$=0.75 mJy/b.}
\label{fig:fig5}
   \end{figure*}
%

\subsection{The radio halo and relic A\,521}
\label{sec:imag521}

In  Fig.  \ref{fig:fig1} (left panel) we show the 153 MHz \gmrt full resolution image 
of the $\sim$15\arcmin$\times$15\arcmin\ area containing the cluster A\,521,   
corresponding to the cluster virial radius (R$_V$=2.78 Mpc; G06).  
The inner circle highlights the radio 
halo region,  $\sim$1 Mpc, and it is the same area as shown in  Fig.  1 of B08. 

The cluster radio emission is dominated by the extended, arc--shaped 
radio relic, located in the South--Eastern periphery of the cluster. 
At this resolution, only part of the central radio halo is visible in the 
$\sim$1 Mpc circular region. 
In order to properly image and analyze the radio halo and relic emission, 
we subtracted from the \textquotedblleft{central}\textquotedblright\ uv--data (Section \ref{sec:obs})  
 the clean components of all the unrelated individual radio sources (compact and extended) 
imaged at full resolution. 
Apart for the bright source J0454-1016a, (studied in 
G08 and labelled as A in  Fig.  \ref{fig:fig1}), 
none of the compact sources detected at 1.4 GHz within either the radio halo 
or the relic were 
detected at 153 MHz (\ie sources B and C in the relic, Fig 2 of G08; D09). 
This is most likely due to the worse surface brightness sensitivity of  
our data. Note that only sources with spectra steeper than 1.5 would be 
detectable at 153 MHz at the sensitivity of our images.
\\
The total flux density of source A at 153 MHz is 65$\pm$13 mJy, 
in agreement (within the uncertainties) with its integrated spectrum in the 
frequency range 327 MHz -- 8 GHz derived in G08. 
To carefully subtract this source, that lies within the relic an the halo,  
the model image was produced cutting the innermost region of the uv-plane, 
to avoid any possible contribution to the flux density from the underlying diffuse emission. 
We then subtracted the relative clean components from the uv--data, ensuring the consistency 
between the total flux density subtracted and the one measured in the full resolution image. 
From the residual dataset we finally produced a low resolution image, by using 
robust 0 weighting and  tapering the visibilities to give more weight to the shortest baselines, 
in order to enhance the large--scale low brightness radio emission.

The final low resolution image, containing only the diffuse cluster emission, 
is presented in right panel of  Fig.  \ref{fig:fig1}. The local noise 
(1$\sigma$) level is 1.45 mJy/b, slightly worse than in the full resolution image, as expected 
as consequence of the resolution and weighting scheme used.  
For comparison with our previous images, the 153 MHz contours are overlaid 
to the \vla image at 1400 MHz, with similar angular resolution of  
$30^{\prime\prime}\times30^{\prime\prime}$ (from D09). 

\subsubsection{Morphological considerations}
\label{sec:morph521}

The low resolution 153 MHz image (Fig. \ref{fig:fig1}, right) clearly 
shows that the radio relic is the brightest feature of the cluster.  
Its overall morphology is in good agreement with the 1400 MHz image.
However, similarly to what observed at 240 MHz (B08),  it 
is slightly more elongated in the S--W edge at 153 MHz than at 1400 MHz, with 
a size of $\sim$ 1.35 Mpc along the major axis. \\
A radio emission, labelled \textquotedblleft spur\textquotedblright in the
left panel of Fig. \ref{fig:fig1}, is visible just South of source A, 
extending westwards from the inner side of the relic.
The \textquotedblleft spur\textquotedblright\ is still visible in the 
subtracted, low-resolution image. A similar structure, partially coincident 
with this feature is also visible in the 240 MHz image (B08), but it is 
not detected in this region at 1400 MHz (Fig. \ref{fig:fig1}, right).  
It is not clear whether the \textquotedblleft spur\textquotedblright it is 
related to the relic, or it is part of the central radio halo. 

The morphology and size of the radio halo at 153 MHz are similar to 
those at 1400 MHz, with the main axis slightly elongated in the N--W direction. 
The similarity between 153 and 1400 MHz is consistent with the fact that 
the two maps have a comparable surface brightness sensitivity: 
assuming a spectral index for the radio halo of $\alpha$=1.8 (D09), and 
extrapolating the image brightness sensitivity from 153 MHz to 1400 MHz, 
we find 
$\simeq$28 $\mu$Jy/b, very close to the sensitivity limit of the 1400 MHz map 
(30$\mu$Jy/b). 
The halo is slightly more extended at 240 MHz than at 153 MHz, with a 
faint region of emission extending toward South not detected at 
153 MHz (see Fig. 1 in B08). 
This is most likely due to the worse surface brightness sensitivity of the 
153 MHz image compared to the 240 MHz one, which was obtained from a much 
deeper ($\sim$ 18 hrs) observation. Indeed, the surface brightness of this 
faint region at 153 MHz, estimated from the 240 MHz image 
assuming a spectral index $\alpha$=1.8 (D09), 
is comparable with the sensitivity limit of our 153 MHz image. 

A striking feature is a \textquotedblleft bridge\textquotedblright of
radio emission connecting the northern part of the relic with the central 
radio halo (Fig. \ref{fig:fig1} right).
The \textquotedblleft bridge\textquotedblright was observed also at 1.4 GHz 
(D09), but it becomes much more prominent at 153 MHz and 240 MHz (B08; 
Fig. \ref{fig:fig1}). As a matter of fact, at these low frequencies this bridge 
becomes so extended that it becomes very difficult to separate the emission 
of the halo from that of the bright relic: large scale diffuse emission 
permeates 
the whole cluster volume, from the inner edge of the relic to the North--West 
of the cluster. 
We note that no point sources are present in the bridge region (except for 
source A, which was subtracted from the u--v plane)
therefore the overall emission connecting the relic and the halo is  
not due to inaccurate or incomplete subtraction of individual sources. 
\\
A number of radio \textquotedblleft bridges\textquotedblright, \ie radio 
emission connecting the halo and relic in the same clusters, are known in the 
literature. Beyond the famous and prototype
\textquotedblleft bridge\textquotedblright in the Coma cluster (Kim et al.
1989), other examples are A\,2744, A\,520, A\,754 (Markevitch 2010; V13).

\subsection{The radio halo in A\,697}
\label{sec:imag697}

In the left panel of  Fig.  \ref{fig:fig3} we present the \gmrt 153 MHz 
full resolution (26.2\arcsec$\times$20.8\arcsec) image of the A\,697 cluster 
field, $\sim$12$^{\prime}$$\times$12$^{\prime}$ wide (corresponding to about 
half the cluster virial radius, R$_V$ = 2.9 Mpc, M10). 
Apart from the two extended radio galaxies S1 and S2 (see 
Fig. \ref{fig:fig3}) and one bright resolved source located 
South--West of the cluster centre, the faint radio halo is the dominant 
feature.\\ 
None of the point sources embedded in the radio halo region at 325 MHz (M10)
has been detected at 153 MHz, not even 
source A, associated with the BGC galaxy at the cluster centre.
Their location is reported in the left panel
of  Fig.  \ref{fig:fig3} (labelled as A, D, G as in Fig 2 of M10). 
\\
Similarly to the case of A\,521 (Sect. \ref{sec:imag521}), from the uv--data 
we subtracted all the discrete radio sources in the field 
and produced a low--resolution image of the radio halo, presented in the
right panel of  Fig.  \ref{fig:fig3}), overlaid on the 325 MHz image at a 
similar resolution (from M10). The largest linear size of the radio halo
is $\sim$1.3 Mpc, and its surface brightness is centrally peaked,
as seen also at higher frequencies (see V08, Van Weeren et al. 2011).
The surface brightness sensitivity of the 325 MHz image is $\sim$2--2.5 better 
than that at 153 MHz (assuming a spectral index of 1.6 for the radio halo, 
\ie Van Weeren et al. 2011), nevertheless, we managed to to detect the 
153 MHz radio halo emission within the region considered for the flux density 
measurements (Fig. \ref{fig:fig3}, right). 
\\
Despite an overall agreement in size and morphology at the various frequencies,
part of the emission imaged at 153 MHz is not visible at higher frequencies. 
This could be due to a very steep spectrum part of the halo, however the 
signal--to--noise ratio is too low here to allow for an accurate study of the 
spectral index distribution in this area.


\subsection{ The complex diffuse emission in A\,1682}
\label{sec:imag1682}

The full resolution ($29.5^{\prime\prime}\times18^{\prime\prime}$) 153 MHz 
image of A\,1682 is presented in  Fig.  \ref{fig:fig5}, overlaid on
a 240 MHz grey scale image at the resolution of 
$18.3^{\prime\prime}\times14^{\prime\prime}$ (Venturi et al. 2009). 
Our 153 MHz observations confirm that the radio emission in this cluster 
is very complex, as already clear from the images at higher frequencies 
(V08, V11 and V13). 

The cluster radio emission is dominated by two main extended sources, the  
\textit{E tail}, a large and elongated tail of emission, connected to the 
central strong radio galaxy (associated to the cluster dominant, see V13), 
and the \textit{N--W ridge}, a bright and elongated source 
extending from the cluster centre to the N--W region of the cluster, which 
diffuses into a fainter, less regular, radio structure.
Moreover, a smaller and fainter \textit{S--E ridge} is located in the outer 
part of the cluster.
Their morphology and brightness distribution are very similar to those at 
240 MHz, despite an overall broader emission of the \emph{N--W ridge}. 

Beyond these sources, the 153 MHz image of A\,1682 shows the presence of a
\textquotedblleft{diffuse component}\textquotedblright\ ,
a region of low surface brightness emission located close to the
S--E ridge in the plane of the sky.
It is coincident with a similar feature detected at 240 MHz, and has a largest 
linear size is $\sim$ 350 kpc. Compared to the images at higher frequencies,
this is the most remarkable feature of the 153 MHz image 
(V11 \& V13).
All four components labelled in Fig. \ref{fig:fig5} are named following the 
same notation as in previous papers (V08, V11, V13).

The flux density values at 153 MHz (measured from the image in 
Fig. \ref{fig:fig5}) and at other frequencies (V13) of all 
the four main components of the radio emission are reported in Table \ref{tab:fluxes1682},
along with their spectral index, computed between 153 and 240 MHz.
The diffuse component has a very steep spectrum, while our flux density 
measurement at 153 MHz leads to a flat spectrum for \emph{S--E ridge}.
This result is very unusual for a diffuse radio
source, and difficult to reconcile with the steepness 
at higher frequencies (V13), and needs further investigation. 
The most likely explanation is missing flux density at 153 MHz (see also Section \ref{sec:diff1682}).
\\

\subsubsection{The diffuse component}
\label{sec:diff1682}

Here we focus our attention on the steep-spectrum diffuse component,  
the most remarkable feature in our new 153 MHz image of the cluster. \\
As clear from Fig. \ref{fig:fig5}, this component 
is already visible at 240 MHz, with S$_{\rm 240~MHz}=46\pm$4 mJy 
(V13, see also Table \ref{tab:fluxes1682}). 
Part of its emission at 240 MHz is undetected at 153 MHz at the 
3$\sigma$ level, and this is likely due to the worse sensitivity of 
this observation (the most affected by data loss of our sample, 
see Sect. \ref{sec:obs}).  
Its total flux density, integrated over the polygonal area shown in
 Fig.  \ref{fig:fig5}, is S$_{153 \rm MHz}=$98$\pm$20 mJy. The resulting spectral 
index in this frequency range is $\alpha^{240 \rm MHz}_{153 \rm MHz}=1.7\pm 0.1$ 
(Table \ref{tab:fluxes1682}). 
This value is flatter than the one between 240 and 610 MHz 
($\alpha$=2.09$\pm$0.15, V13).  
This might be due to the fact that we are missing part of its flux density at 
153 MHz in the integrated area, due to 
the worst surface brightness sensitivity at 153 MHz (estimated to be about a 
factor of 3 lower than that at 240 MHz, assuming a spectral index of 1.7). 
Alternatively, the spectral index between 240 and 610 MHz might be flatter 
due to possible flux density losses at 610 MHz.
Despite the uncertainty in the spectral steepness, arising from the different 
sensitivities of the observations, we can safely conclude that the diffuse 
component in A\,1682 has a very steep spectrum. 

Beyond the main four components, excess of radio emission spread over the 
cluster scale (Mpc--size) both at 240 and 610 MHz was reported in V08, 
V11 and V13, estimated by subtraction of the individual 
sources (\emph{N--W ridge}, \emph{E tail}, \emph{S--E ridge}). 
Hints of large scale emission were also found from re-analysis of VLA 
D--array data at 1.4 GHz (see V11). 
This suggests the presence of faint, diffuse very low brightness emission.\\
In search for any evidence of this at 153 MHz, we subtracted the flux density 
of the \emph{N--W ridge}, \emph{E tail} and \emph{S--E ridge} from the total 
integrated flux density over a $\sim$1 Mpc area, and we found   
a residual flux density $\sim250$ mJy at 153 MHz, and $\sim$160 mJy at 240 MHz, 
consistent with a spectral slope of 1 (as found from the residual analysis 
in V11, and confirmed at other frequencies in V13). 
Considering the large uncertainties at all frequencies, from these
results it is difficult to draw any conclusion on this 
diffuse cluster scale residual emission and its nature. \\


\section{Spectral analysis}
\label{sec:analysis}

The high sensitivity 153 MHz \gmrt observations presented here allow us to 
study the spectral properties of the radio halos in A\,521 and A\,697, and 
of the relic in A\,521 over more than one order of magnitude.

\subsection{The integrated spectra}

The integrated spectra of radio halos and relics are a crucial piece of 
information for our understanding of their origin, since they bear the 
signature of the physical processes driving their formation. 
So far, only a handful of integrated 
spectra are well sampled down to frequencies of the order of 100 MHz, and 
below (Coma, Thierbach et al. 2003; A\,1914, Bacchi et al. 2003; A\,2256, 
Brentjens 2008, Van Weeren et al. 2012), but even in those cases, the 
resolution of the available observations is too low to separate the diffuse 
Mpc--scale emission from that of the embedded sources. Moreover, matching
the images obtained with different instruments, techniques and resolutions
introduces further uncertainties which may seriously bias the results.
One of the most outstanding features of the GMRT is the capability to 
perform simultaneous imaging at both high and low resolution, thanks to 
its antenna configuration. This allows to isolate the emission of the halos 
and relics from the embedded individual sources with high accuracy.

For A\,521 and A\,697 we have been able to perform accurate and consistent
subtraction of the discrete embedded sources at all frequencies at which they 
give a contribution to the total flux density measurements, and we have 
obtained integrated spectra of the ``pure'' diffuse emission down to 153 MHz.\\

\subsection{The relic and halo in A\,521}

A detailed spectral analysis of the relic in A\,521 in the frequency range
240 MHz -- 5 GHz was performed in G08, while the integrated spectrum of the 
radio halo in A\,521 in the frequency range 240 MHz--1.4 GHz was studied in 
D09. 

\begin{figure}
   \centering
\includegraphics[angle=0,scale=0.45]{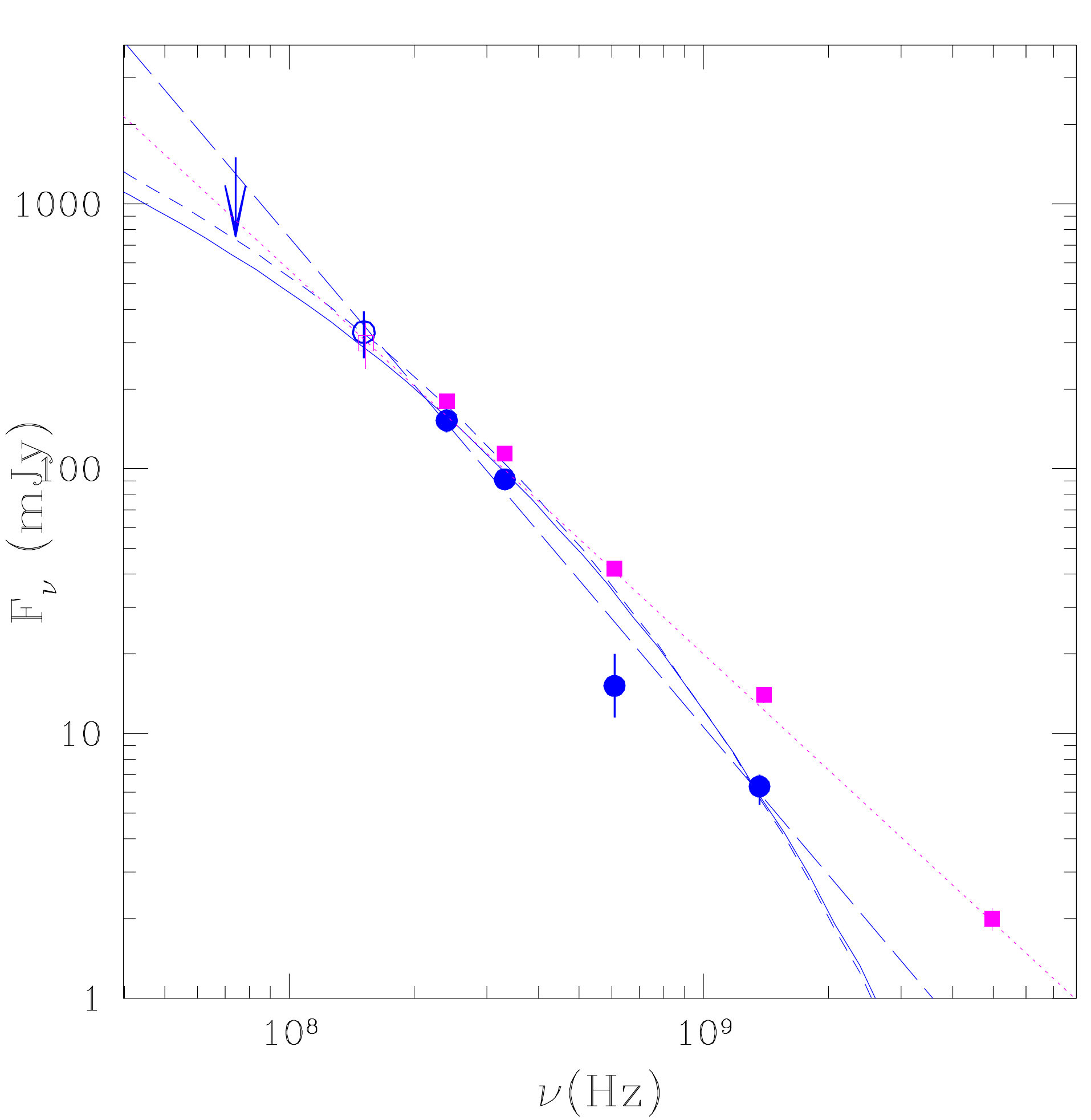}
      \caption{Integrated spectra of the halo and relic sources in A\,521. 
       Radio halo spectrum (blue):  the empty circle is the \gmrt 153 MHz flux 
(this work), filled circles are measurements from B08 and D09; the 
power--law fit to the data weighted for the uncertainties (dashed blue line)
corresponds to $\alpha$=1.85.
Radio relic spectrum (magenta): filled squares are the flux density values
taken from G08, the empty square is the \gmrt 153 MHz flux density (this work);
the magenta dashed line is the power law fit (weighted for the uncertainties) 
corresponding to $\alpha$=1.45.
The curved dashed and solid blue lines are the fit of homogeneous 
re--acceleration 
models, with constant acceleration efficiency in the radio emitting volume 
(constant ratio of turbulent and thermal energy densities). 
The magnetic field is assumed to scale with radius (azimuthally constant) as 
 $\rho^{2/3}$ (short-dashed) and as $\rho$ (solid); flatter radial
 distributions of the magnetic field strength produce more synchrotron 
 power at lower frequencies.
}
\label{fig:fig2zoo}
   \end{figure}

In order to obtain a 153 MHz flux density measurement consistent with the 
previous studies, we integrated the total flux density of the relic over 
the same region used in G08, and we derived the total flux density of the 
radio halo by integrating the image in Fig. \ref{fig:fig1} (right) over 
the same circular area of 1 Mpc used in B08 and D09. 
We found 
S$_{\rm 153~MHz}$(relic)=297$\pm$59 mJy and 
S$_{\rm 153~MHz}$(halo)=328$\pm$66 mJy 
(see Table \ref{tab:fluxes}).
We estimate that any possible contribution of point sources embedded in the 
relic and halo (not detected at this frequency, see Sect. \ref{sec:imag521}) 
to the total 153 MHz flux density measurement is $\lax$2.3  mJy,
well within the uncertainties of the measured flux densities. 
This estimate was obtained by extrapolating the total contribution of these 
sources at 1400 MHz ($\lax$0.4 mJy), down to 153 MHz, assuming an 
average spectral index for each source of 0.7--0.8 (see D09). \\
The updated integrated spectra of the radio halo and of the relic 
are shown in Fig. \ref{fig:fig2zoo} (blue circles and magenta squares, 
respectively). Filled symbols are taken from D09, while the empty ones
are the new 153 MHz measurement. 

We fitted both spectra with a power--law. As clear from Fig. \ref{fig:fig2zoo}
the best fit of the halo spectrum is dependent on the inclusion of the flux 
density value at 610 MHz. We obtain $\alpha_{\rm 153~MHz}^{\rm 1.4~GHz}=1.91\pm0.11$ 
and $\alpha_{\rm 153~MHz}^{\rm 1.4~GHz}=1.81\pm0.02$, respectively with and without 
it. 
The spectrum of the relic is fitted with a power--law with spectral index
$\alpha_{\rm 153~MHz}^{\rm 5~GHz}=1.45\pm0.02$.
\\
The inclusion of the flux density at 153 MHz confirms the previous values 
of the spectral index both for the halo (D09) and for the relic (G08).

The spectrum of the radio halo was compared to a homogeneous turbulent 
re--acceleration model (using Brunetti \& Lazarian 2007), reported in Fig.
\ref{fig:fig2zoo} as solid and short--dashed curved blue lines. 
For the models we have
assumed that {\it (1)} the acceleration efficiency is constant over the whole 
radio emitting volume, with constant ratio of turbulent and thermal energy 
density, and {\it (2)} the magnetic field strength declines with increasing 
radius (azimuthally constant) 
as $\rho^{2/3}$ and as $\rho$ (dashed and solid curved line respectively, $\rho$ being the density of thermal protons); 
the first model  assumes the magnetic field is frozen into the thermal gas, while the 
second model is expected by some numerical simulations (\eg  Dolag et al 2005). 
We note that some observations preferentially derive 
a flatter magnetic field radial distributions, 
\ie $B \propto \rho^{1/2}$, 
expected in the case in which the magnetic energy density scales  
as the thermal energy density (assuming a constant gas temperature) 
 (\eg Coma cluster, Bonafede et al. 2010). \\
The long dashed blue line is a power--law with 
$\alpha=1.85$, the intermediate steepness between the two fits with and 
without the inclusion of the 610 MHz flux density value.

%

  \begin{figure}
   \centering
\includegraphics[angle=0,scale=0.45]{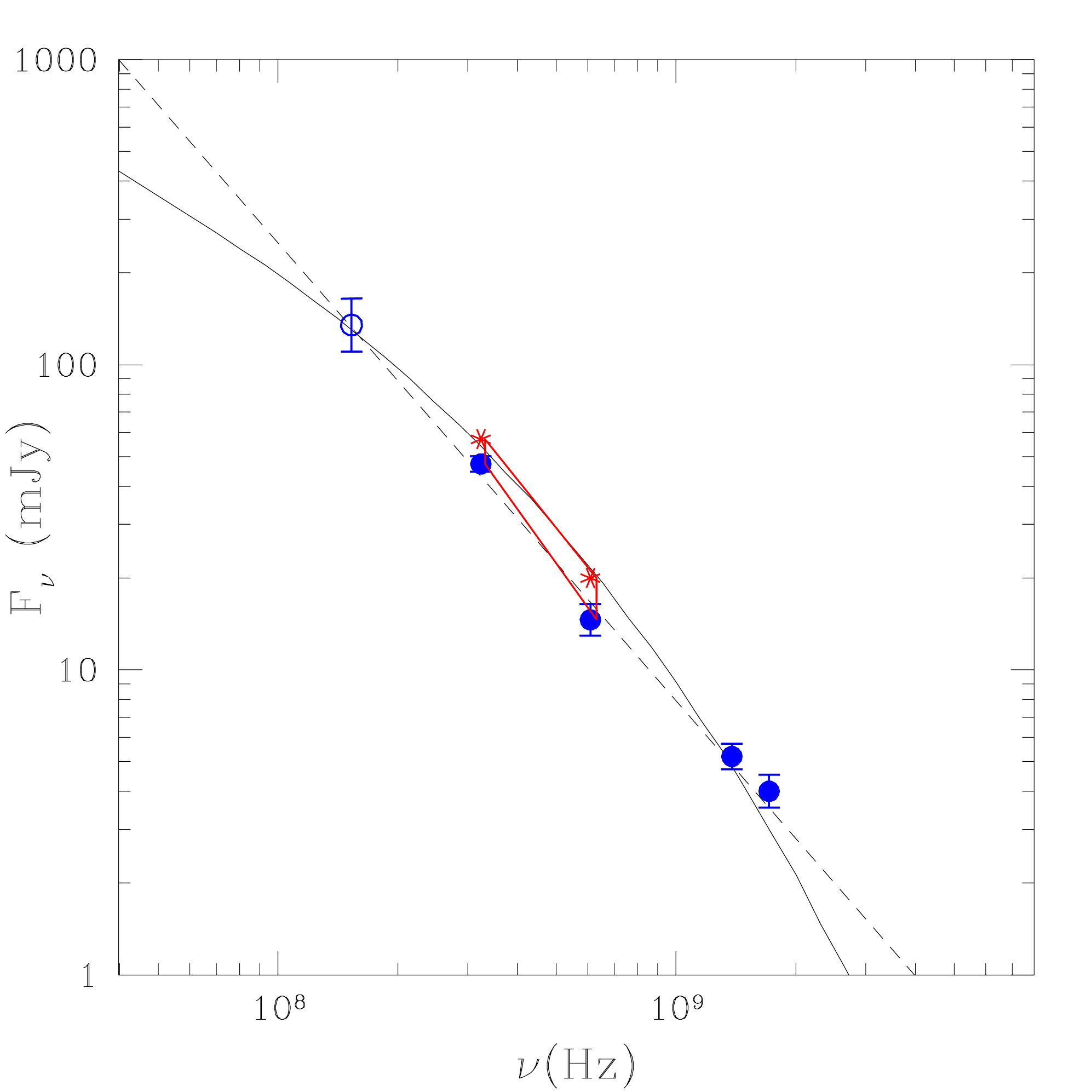}
      \caption{Integrated spectrum of the radio halo in A\,697. 
The blue empty circle is the \gmrt 153 MHz measurement derived in this paper.
\gmrt flux density measurements and limits at 330 and 610 MHz (filled 
circles and red asterisks respectively ) are taken from M10, while the 
WSRT flux densities at 1382 and 1714 MHz are taken from van Weeren et al. 
2011. The dashed line is the power--law fit to the data, weighted for the 
uncertainties ($\alpha$=1.5); the solid curved line is from the model 
of homogeneous re--acceleration, with constant acceleration efficiency 
in the radio emitting volume (constant ratio of turbulent and thermal 
energy densities). The magnetic field is assumed to decline with radius 
(azimuthally constant) as $\rho$. 
The (red) butterfly and asterisks mark the fluxes at 325 and 610 MHz
corrected for the missing flux (M10).}
\label{fig:fig4}
   \end{figure}
%

\subsection{The radio halo in A\,697}
\label{sec:spec697}

For a consistent comparison with the flux density measurements in the 
frequency range 325 MHz--1.4 GHz (M10) we integrated the low resolution 
153 MHz image over the same circular area of $\sim$1.3 Mpc used for the 
flux measurements at higher frequencies (shown by the dashed circle in  Fig.  
\ref{fig:fig3}, right panel). We find S$_{153\rm{MHz}} =$135$\pm$27 mJy. 
We estimate that the upper limit to the contribution to the total radio
halo flux density of the point sources embedded in the halo (A to G, not 
detected at 153 MHz, see Sect. \ref{sec:imag697}) is $\lax$18 mJy. This value
was obtained extrapolating their flux density from 325 MHz assuming an 
average spectral index for the sources $\lax$1 
(as derived from higher frequency measurements), 
and it is within the uncertainties to the flux density measurements. 

The updated integrated spectrum of the halo is shown in Fig. \ref{fig:fig4}. 
The empty circle is the new 153 MHz value; filled circles and stars at
325 MHz and 610 MHz are measurements and limits from M10 (see the paper for 
more details), the values at 1382 and 1710 MHz are the recent WSRT 
measurements by van Weeren et al. 2011 (see Table \ref{tab:fluxes}  for the values). \\
The data points can be fitted with a single power--law spectrum down to 153 
MHz, with a slope which is only marginally affected by the results of the
detailed analysis on the missing flux density performed in M10.
In particular, $\alpha_{\rm 153~MHz}^{\rm 1.4~GHz} = $1.50$\pm$ 0.03
and $\alpha_{\rm 153~MHz}^{\rm 1.4~GHz} = $1.52$\pm$ 0.05 for the observed (blue
circles) and corrected (red stars) GMRT values respectively (M10).
This value is slightly flatter than reported in van Weeren et al. 2011 in the
frequency range 325 MHz--1.7 GHz, as our 153 MHz flux density measurement
is below the extrapolation of their power--law.

The shape and steepness of the spectrum of the radio halo in A\,697 were 
compared to a homogeneous turbulent re--acceleration model 
(Brunetti \& Lazarian 2007). The result is plotted in Fig. \ref{fig:fig4},
for a magnetic field intensity scaling as $\rho$.
The dashed line is a power--law with spectral index $\alpha$=1.5.

%
\begin{table*}[ht!]
\caption{Flux density and spectral index of the components in A\,1682.}
\begin{tabular}{|c|cccc|c|}
\hline
\multicolumn{1}{|c|}{$\nu$ [MHz] }& \multicolumn{3}{r}{$S(\nu)$ [mJy]} &  \multicolumn{1}{c|}{} & \multicolumn{1}{c|}{Ref.}\\
\hline \hline
\multicolumn{1}{|c|}{ }& \multicolumn{1}{c|}{N-W ridge} & \multicolumn{1}{c|}{S-E Ridge}  
& \multicolumn{1}{c}{E Tail} & \multicolumn{1}{|c|}{Diff. Comp.} &   \multicolumn{1}{c|}{} \\ 
\hline
153 	& 746 $\pm$149 &  62$\pm$12  &   1710$\pm$342  &  98$\pm$20 & This work \\
240 	& 468$\pm$15    &  63$\pm$4 	  &  1226$\pm$40     &  46$\pm$4     & V13\\
\hline\hline
$\alpha^{240 \rm MHz}_{153 \rm MHz}$ 	& 1.1$\pm$0.1 	& --$^{(a)}$ &   0.7$\pm$0.1 & 1.7$\pm$0.1  & This work\\
\hline
\end{tabular}
\\
$^{(a)}$ See Section 4.3
\label{tab:fluxes1682}
\end{table*}

%
%
\begin{table}[ht!]
\begin{center} 
\caption{Flux density and spectral index of the cluster sources in A\,521 and A\,697.}
\label{tab:fluxes}
\begin{tabular}{|c|cc|c|c|}
\hline
\multicolumn{1}{|c|}{$\nu$ [MHz] }& \multicolumn{2}{r}{$S(\nu)$ [mJy]} &  \multicolumn{1}{c|}{} & \multicolumn{1}{c|}{Ref.}\\
\hline\hline
\multicolumn{1}{|c|}{ }& \multicolumn{2}{c|}{A\,521} & \multicolumn{1}{c|}{ A\,697} & \multicolumn{1}{c|}{} \\
\hline 
\multicolumn{1}{|c|}{ }& \multicolumn{1}{c}{Relic} & \multicolumn{1}{c|}{Halo}  & \multicolumn{1}{c|}{ Halo} & \multicolumn{1}{c|}{} \\ 
\hline
153 	& 297$\pm$59 &  328$\pm$66  & 135$\pm$27	& This work\\
74		& 660 			& 1500.0 		& --			& G08; B08\\ 
235  	& 180$\pm$10  & --				& --			& G08\\
240   	&  --			& 152$\pm$15  & --  			& B08\\
325    	&  --			&  90$\pm$7	& 47.3$\pm$2.7 & B08; M10 \\
327    	&  114$\pm$6	&  --	 	 	 & --			  & G08\\
610		&   42$\pm$2	& 15.0$\pm$3.5 & 14.6$\pm$1.7 & G08; B08; M10\\
1365    	&  --			&  6.4$\pm$0.6	 &  -- 			& D09\\
1382    	&  --			&  --	 		 &  5.2$\pm$0.5  & VW11\\
1400    	&   14$\pm$1 	&  -- 			 & --			& G08\\
1714    	&  --			&  --  	 		 & 4.0$\pm$0.5 & VW11\\
4980    	&  2.0$\pm$0.2 &  -- 		 	& --			 & G08\\
\hline\hline
$\alpha_{fit}$ & 1.45 $\pm$0.02 & 1.81$\pm0.02^{(a)}$ & 1.52$\pm$0.05 &   \\
\hline
\end{tabular}
\end{center}
$^{(a)}$ Value obtained without the inclusion of S$_{\rm 610~MHz}$ (see Sect. 5.2).
\end{table}

\section{Discussion}
\label{sec:discussion}

In this paper we have presented high sensitivity and high resolution 
GMRT observations at 153 MHz of the very low surface brightness diffuse 
emission in the galaxy clusters A\,521, A\,697 and A\,1682, belonging
to the GMRT Radio Halo Cluster Sample, and extensively studied over
a very wide range of frequencies, from 153 MHz to 1.4 GHz. 
Despite the massive data editing, as consequence of the strong RFI at this
frequency, the quality of our final images is very good, with 1$\sigma$
noise level in the range 0.8--1.7 mJy b$^{-1}$.

Our observations almost double the spectral information available to
date for cluster radio halos and relics in a range of frequencies 
which is critical for our understanding of their origin and evolution. 

\subsection{The relic source in A\,521}
\label{sec:RELIC}

The spectrum of the relic in A\,521 is one of the best sampled in the
literature, with 6 flux density measurements in the frequency range 
153 MHz--5 GHz. 
The integrated spectrum of the relic down to 153 MHz is consistent with 
a power--law fit with slope $\alpha=1.45$ (Fig. \ref{fig:fig2zoo}),  
confirming the previous results (G08, B08). 
G08 discussed in detail the possible models for its origin, and concluded 
that the most likely is the Fermi-I diffusive shock acceleration 
scenario, in which the relativistic electrons responsible for the radio 
emission are directly accelerated from the thermal gas due to the passage of 
a merger shock wave (\eg Ensslin et al. 1998; Hoeft \& Br\"uggen 2007).  
Our result shows that a power--law spectrum still holds down to 153 MHz,
further supporting the diffusive shock acceleration scenario.
Moreover, the presence of a shock front in the X--ray gas at the location 
of the relic has been recently revealed by XMM analysis of the cluster, as a 
both a density and temperature jump (Bourdin et al. 2012).
  
In the shock acceleration model the Mach number of the shock responsible for 
the acceleration of the electrons is directly related to the observed slope 
of the 
power--law spectrum: for $\alpha=1.45$, the corresponding Mach number of the 
shock is $\sim$2.3 (see G08 for more details). This is consistent with the 
Mach number estimated from the X-ray density jump, $M= 2.4\pm0.2$ 
(Bourdin et al. 2012). We note that the single power--law spectrum poses
a new challenge, since it requires that a weak shock, as the one in
A\,521, is able to channel a substantial fraction of the shock energy into
acceleration of the relativistic electrons (Kang et al. 2012).

\subsection{Origin of the very steep spectrum radio halos in A\,521 and A\,697}
\label{sec:HALOS}

Figs. \ref{fig:fig2zoo} and \ref{fig:fig4} show the observed spectrum of the 
radio halo in A\,521 and A\,697 respectively, with overlaid the best fit 
power--law (straight dashed line in both cases) and the 
turbulent re--acceleration model (curved lines). 
Our data confirm that the power--law spectrum with the steepness derived in 
previous works (B08 and D08 for A\,521; M10 and van Weeren et al. 2011
for A\,697) still holds going to 153 MHz. In both cases it is impossible 
to distinguish between a simple power--law shape and a curved shape as 
expected by homogeneous re--acceleration models. In particular:

\begin{itemize}
\item[{\it (i)}]
The very steep power--law spectrum of the A\,521 radio halo, with 
$\alpha_{\rm 153~MHz}^{\rm 1.4~GHz}$ in the range 1.8--1.9, implies that the 
spectrum 
of the emitting electrons be extremely steep, \ie $\delta_{inj} \sim 3.8$.
Based on simple energetic arguments, we expect that the spectrum
flattens at very low frequencies. 
As a matter of fact, the extrapolation of such steep
power--law  spectrum down to trans--relativistic energies would imply an 
uncomfortably large energy budget in the form of relativistic electrons;
the situation becomes even more severe if we extrapolate such power--law 
spectrum to supra-thermal energies. The case of A\,521 is similar to
the Coma radio halo spectrum, which is straight if we restrict our analysis 
to the $\sim$ 600 MHz -- 5 GHz frequency range, while the curvature becomes
clear with the inclusion of the flux density measurements at lower 
frequencies (Thierbach et al. 2003).\\
The upper limit at 74 MHz, derived from the VLSS, is very close to the 
power--law extrapolation with $\alpha=1.85$, therefore we trust that future
observations at $\leq 100$ MHz with the LOw Frequency ARray (LOFAR) will 
allow a direct test of the presence of a spectral curvature.

\item[{\it (ii)}]
In the case of A\,697, the steepness of the best fit power--law, 
$\alpha_{\rm 153~MHz}^{\rm 1.4~GHz}$ =1.5 implies that the power law of the 
emitting electrons is $\delta_{inj} \sim 3$. 
Even this value, extrapolated to very low frequencies (energies) implies
that the energy budget in the form of relativistic electrons becomes very 
large.
This would lead to a contradictory case, where the acceleration mechanism 
is poorly efficient (as suggested by the steep spectrum) but, at the same 
time, it is able to extract a substantial fraction of the energy in the ICM
in the form of cosmic rays.
Even for this cluster, future observations at the frequencies offered by 
LOFAR will allow to test the presence of a spectral curvature.

\end{itemize}

The curved spectra shown in Figs. \ref{fig:fig2zoo} and \ref{fig:fig4}
have been derived assuming  homogeneous models. However, the recent 
LOFAR observations of A\,2256 (van Weeren et al. 2012) suggest that 
the situation may be more complex, at least for that cluster.
As a matter of fact, turbulence can be intermittent and not homogeneous,
and the ratio between turbulent and thermal energy density may
change with space and time in the radio--emitting volume.
Under that hypothesis, the spectrum of the emitting electrons would 
actually be a mix of different populations of accelerated particles, 
and as a result of this the synchrotron spectrum, integrated along 
the line of sight would be more complex than those reported in
Figs. \ref{fig:fig2zoo} and \ref{fig:fig4}.
\\
In addition, homogeneous models assume an azimuthally--averaged value 
of the magnetic field in the radio halo, while a better representation
of its properties should take into account point--to--point variations 
(or scatter) of the field intensity across such averaged value.
If the magnetic field is inhomogeneous, the resulting synchrotron
spectrum would be stretched in frequency, making any spectral curvature
even smoother and more difficult to observe.

Despite the difficulty to disentangle between different models and to 
provide detailed theoretical predictions in the case of inhomogeneous media, 
the most relevant fact arising from our observations is the confirmation 
of the very steep spectrum of these two radio halos.
For a given radio luminosity at 1.4 GHz the energy budget required in the 
form of relativistic electrons in a halo with spectral index 
$\alpha\sim1.85$ (A\,521) is about $10^4$ times larger than that required in 
a halo with spectrum $\alpha=1.2$, provided that the spectral energy 
distribution of relativistic electrons can be extrapolated to lower 
(trans--relativistic) energies assuming a power--law with injection slope 
$\delta_{inj} = 2 \alpha$.

This analysis highlights the serious difficulty to explain the origin of
ultra--steep spectrum halos, and motivates the need for a flattening of 
the relativistic electrons (and synchrotron) spectrum at low energies 
(frequencies).

\subsection{A hidden radio halo in A\,1682?} 
\label{sec:disc1682}

Our 153 MHz observations further confirm that the radio emission at the 
centre of A\,1682 is very complex, and difficult to fit into our current
knowledge of diffuse cluster scale emission.

\begin{figure}
   \centering
\includegraphics[angle=0,scale=0.4]{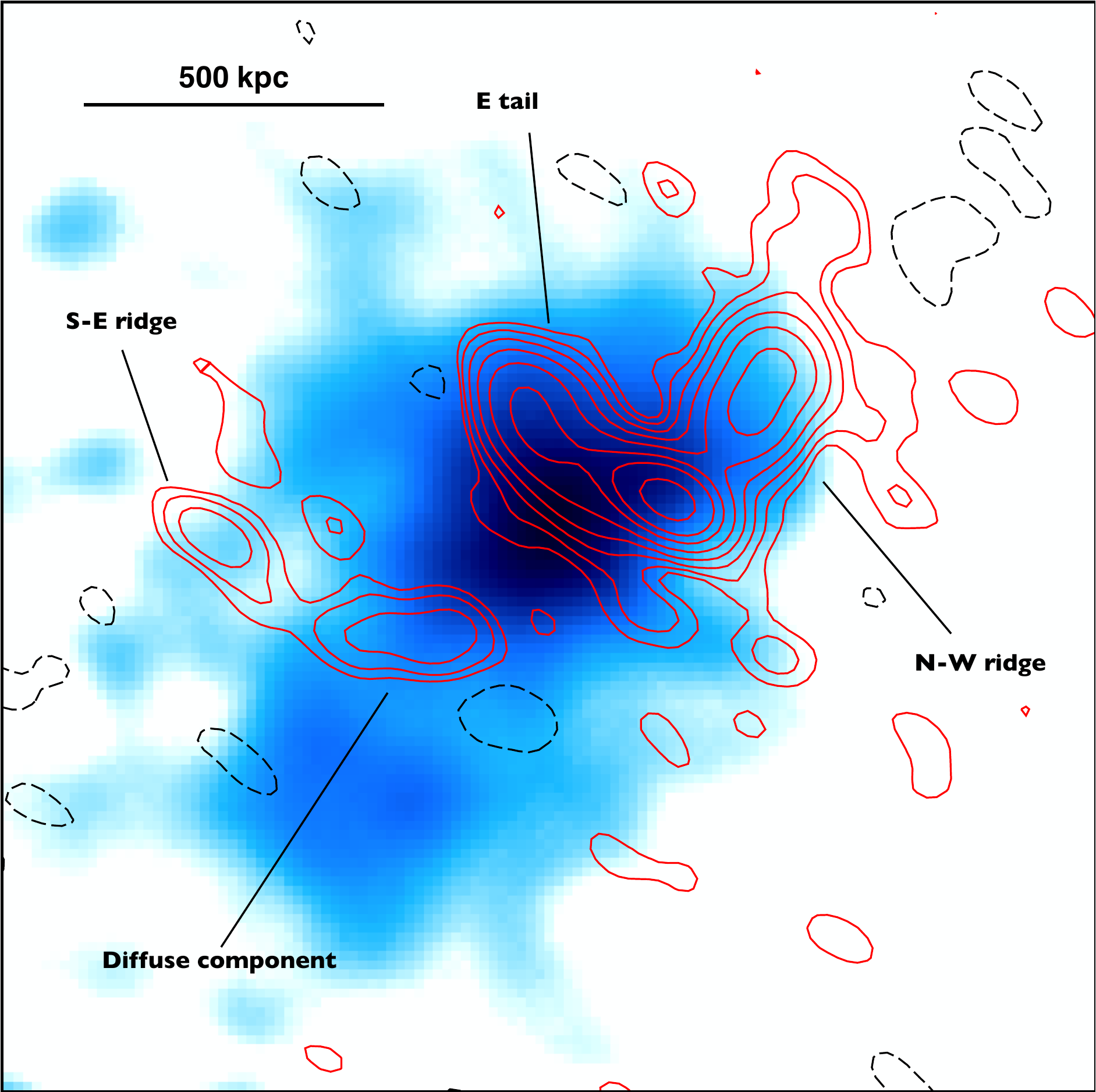}
      \caption{ \gmrt 153 MHz radio contours of A\,1682 (same as in 
Fig. \ref{fig:fig5}), overlaid to the archival \chandra\ smoothed X--ray 
image.}
\label{fig:fig7}
   \end{figure}

At 153 MHz all the four main components of emission visible at higher
frequencies have been clearly identified (Fig. \ref{fig:fig5}, 
Table \ref{tab:fluxes1682}). 
For a qualitative comparison of the non--thermal and thermal properties of 
this complex merging cluster, in Fig. \ref{fig:fig7} we show the 153 MHz 
radio emission of the cluster centre overlaid on the \chandra\ X--ray 
smoothed image (taken from archive data, Obs. Id. 3244, ACIS--I, exposure 
10 ks, no analysis was performed). 
The gas distribution in A\,1682 is clearly disturbed, with two main 
condensations, the North--Western being considerably brighter and more 
extended than the South--Eastern one. 
We note a clear offset between the radio galaxy 
associated with the BCG and the peak of the X-ray cluster emission, which 
is a further hint of the disturbed dynamical status of the system.  
The bright part of the N--W ridge is located just outside the central 
brightest X--ray region, while the fainter radio extension to the 
West/North-West lies in a more peripheral area, at the edge of the cluster 
X--ray emission.   
The S--E ridge, too, is located in a peripheral region of very faint X--ray 
emission. 
The steep spectrum diffuse component (see Section 4.3.1) lies in 
between the two condensations, in a region of decreased X--ray surface 
brightness. 

Despite the multifrequency radio observations available, the nature of each 
component remains enigmatic. 
A recent spectral analysis of the E--tail (V13) shows that 
its connection with the cluster BCG is unlikely, while the elongated shape, 
size ($\sim$ 600 kpc), spectrum (Table \ref{tab:fluxes1682}) and peripheral 
location (Fig. \ref{fig:fig7}) of the N--W ridge are suggestive of a radio 
relic (see also V13).

The most intriguing feature is the diffuse component. Our data confirm its 
very steep spectrum ($\alpha\lax$1.7; see also Table \ref{tab:fluxes1682}). 
One possible explanation for the nature of this structure is that we 
are looking at the brightest region of an underlying, very low surface 
brightness, giant radio halo, whose presence is suggested by the presence 
of positive residual of emission over a scale of $\sim$ 1 Mpc, detected at 
153 MHz, 240 MHz and 610 MHz (this work, V08, V13, see
Sect. 4.3). 
An alternative possibility is that this component is associated to a dying 
radio galaxy. The elongated morphology imaged at 153 MHz, coupled with the
location at the border of the X--ray emission, are also suggestive of a 
relic source, however this interpretation is difficult to reconcile with
the detections at higher frequencies.

A\,1682 was recently observed with LOFAR (August 2011), as part of the telescope
commissioning observations in LBA (Low Band Array; 15--77 MHz);  
preliminary images around 60 MHz confirm the complex emission in this cluster 
and, in particular, the presence of the steep spectrum \emph{diffuse component} (Macario et al. in prep.). 

%

\section{Summary and conclusions}
\label{sec:summary}

We have presented GMRT 153 MHz high sensitivity imaging of the diffuse cluster 
scale emission in the three galaxy clusters A\,521, A\,697 and A\,1682.
Despite the massive presence of RFI at this frequency, resulting in a
considerable fraction of data removal during the editing process, our images
are of very good quality, with 1$\sigma$ noise level in the range 0.8--1.7
mJy b$^{-1}$ in the full resolution images.
All clusters belong to the GMRT Radio Halo Cluster Sample (Venturi et al. 2007, V08), and 
have been observed with the GMRT also at 240, 325 and 610 MHz (B08, G08, M10,
V11 and V13) and with the VLA at 1.4 GHz (D09, G08). Our 
spectral coverage thus covers about one order of magnitude in frequency.

The morphology and brightness distribution of the radio halo and relic in
A\,521 are similar to those at higher frequencies, and its size is similar 
to what has been imaged at 1.4 GHz (D09).
Part of the halo was missed at this frequency, compared to the 240 MHz
images (B08), most likely as consequence of the worse quality of the 153 MHz
images. At this frequency, the halo and the relic are connected by a bridge
of emission, as it has been found for a number of clusters observed at 
frequencies $\lesssim$ 325 MHz (V13, Markevitch 2010).
\\
The radio halo in A\,697 is centrally peaked at 153 MHz, and its brightness 
distribution decreases radially. At the sensitivity level of our 
observations, the size of the radio halo is the same as detected at 325 MHz,
\ie $\sim$1.3 Mpc (M10). 
\\
Our observations confirm the presence of the {\it diffuse component} in
A\,1682, a steep spectrum region of emission located South--East of the
cluster centre, and detected just above the sensitivity level of the GMRT 
observations at 610 MHz and 240 MHz (V08, V11). 
Its classification and origin is still under investigation, as well as 
the whole central radio emission in A\,1682. 
Our observations confirm that each component has an overall steep spectrum.

We derived the integrated spectrum of the radio halos in A\,521 and A\,697,
and of relic in A\,521. 
In each case our analysis is based on consistent flux density
measurements, derived after careful subtraction of the embedded sources at all
frequencies.  Our spectra are among the most accurate derived so far
for diffuse cluster scale emission.
The previous findings of very steep spectrum sources have been confirmed 
by our 153 MHz observations. 
The spectrum of the relic in A\,521 is a power--law with slope 
$\alpha_{\rm 153~MHz}^{\rm 5~GHz}=1.48\pm0.02$. 
The very steep spectrum radio halo in A\,521 and A\,697 holds at least 
down to 153 MHz. In particular, 
for the halo A\,521 the slope of the spectrum is in the range 
$\alpha_{\rm 153~MHz}^{\rm 1.4~GHz}=1.8-1.9$ depending on the inclusion of the
610 MHz data point (which is known to suffer from major uncertainties),
while the spectral index we derived for A\,697 with the addition of the
153 MHz flux density measurement is 
$\alpha_{\rm 153~MHz}^{\rm 1.4~GHz}=1.52\pm0.05$, slightly flatter than 
the previous measurements (M10, van Weeren et al. 2011), though all values
are still consistent with each others within the errors. 

Steep spectrum radio halos provide important clues on the origin of
the radiating electrons. On one hand, their steep spectrum poses severe 
challenges to the secondary models (B08, M10), on the other, the
re--acceleration scenario requires that we are sampling the steep region 
of the spectrum, before the low frequency flattening takes place.

For the giant radio halos in A\,521 and A\,697 we compared their integrated 
spectra with the models derived assuming homogeneous turbulent re--acceleration.
In both cases 
the frequency coverage from 153 MHz to 1.4 GHz does not allow to discriminate
between possible models, since under those assumptions, a low frequency 
spectral flattening should become clearly visible only below 100 MHz. 
The very steep spectrum of both halos argues
in favor of a flattening on the basis of the very large energy budget
required if the single power--law approximation were extended at much lower
frequencies. One weakness of our analysis might be the assumption of
homogeneous models, however at this stage it is difficult to provide 
details in case of inhomogeneity. The forthcoming LOFAR high sensitivity
observations below 100 MHz will allow us to test the presence of a
low frequency flattening of the spectrum of radio halos, thus providing
conclusive clues on the problem of their origin.

\begin{acknowledgements}

We thank the anonymous referee for useful comments. 
We warmly thank N. Khantaria for her help with the 
observations and invaluable suggestions during  the data reduction. 
We thank the staff of the \gmrt for their help during the observations. 
\gmrt is run by the National Centre for Radio Astrophysics of the Tata 
Institute of Fundamental Research. 
This work has been partially supported by contract PRIN INAF 2008. 
GM and CF acknowledge financial support by the 
\textquotedblleft{Agence Nationale 
de la Recherche}\textquotedblright\ through grant ANR-09-JCJC-0001-01. 
HTI acknowledges financial support through a Jansky Fellowship 
of the National Radio Astronomy Observatory, which 
is operated by Associated Universities, Inc., 
under cooperative agreement with the National Science Foundation. 
SG acknowledges the support of NASA through Einstein Postdoctoral
Fellowship PF0-110071 awarded by the Chandra X-ray Center (CXC), which is 
operated by the Smithsonian Astrophysical Observatory (SAO).

\end{acknowledgements}

\bibliographystyle{aa}

\end{document}